\documentclass[12pt]{article}
\usepackage{graphicx}
\usepackage{amssymb}
\usepackage{amsmath}

\newcommand{\bear}{\begin{array}}  
\newcommand {\eear}{\end{array}}
\newcommand{\bea}{\begin{eqnarray}}   
\newcommand{\eea}{\end{eqnarray}}
\newcommand{\beq}{\begin{eqnarray}}   
\newcommand{\eeq}{\end{eqnarray}}
\newcommand{\bef}{\begin{figure}}  \newcommand 
{\eef}{\end{figure}}
\newcommand{\bec}{\begin{center}}  \newcommand 
{\eec}{\end{center}}

\newcommand{\oneui}{\overline{\tilde{\chi}^0}}
\newcommand{\sla}[1]{\not\!#1}

\begin{document}

\begin{titlepage}

\begin{flushright}
IPMU~10-0066\\
ICRR-Report-568-2010-1
\end{flushright}

\vskip 1.35cm

\begin{center}

  {\large A complete calculation for direct detection of Wino dark
    matter}

\vskip 1.2cm

Junji Hisano$^{a,b}$, Koji Ishiwata$^a$, and Natsumi Nagata$^a$

\vskip 0.4cm

{ \it $^a$Institute for Cosmic Ray Research,
University of Tokyo, Kashiwa 277-8582, Japan}\\
{\it $^b$Institute for the Physics and Mathematics of the Universe,
University of Tokyo, Kashiwa 277-8568, Japan}

\date{\today}

\begin{abstract} 
  In the anomaly-mediated supersymmetry (SUSY) breaking scenario,
  neutral gaugino of $SU(2)_L$ multiplet, Wino, can be the lightest
  SUSY particle and become a candidate for dark matter.  We calculated
  scattering cross section of Wino dark matter with nucleon, which is
  responsible for direct detection of the dark matter, on the
  assumption that the SUSY particles and the heavier Higgs bosons have
  masses of the order of the gravitino mass in the SUSY standard
  model. In such a case, the Wino-nucleon coupling is generated by
  loop processes.  We have included two-loop contribution to
  Wino-gluon interaction in the calculation, since it is one of the
  leading contributions to the Wino-nucleon coupling. It was found
  that the spin-independent scattering cross section with proton is
  $10^{-(46-48)}~{\rm cm^2 }$.  While it is almost independent of the
  Wino mass, the result is quite sensitive to the Higgs boson mass due
  to the accidental cancellation.

\end{abstract}



\end{center}
\end{titlepage}

\section{Introduction}

The anomaly mediation \cite{Randall:1998uk} is the most economical
mechanism to generate supersymmetry (SUSY) breaking terms in the
supersymmetric standard model (SUSY SM). In the breaking mechanism,
only dynamical SUSY-breaking sector is required, and no other extra
fields are needed. On the assumption of the generic form of K\"ahler
potential, all the scalar bosons except the lightest Higgs boson
acquire masses, which are of the order of the gravitino mass. The
gaugino masses, on the other hand, are generated by the quantum
effects, and then they are suppressed by the one-loop factor compared
with the gravitino mass. This is a concrete realization of the split
SUSY scenario \cite{split}, in which the squarks and sleptons are
$O(10^{(1-2)})$ TeV while the gaugino masses are less than $O(1)$ TeV.
Such a mass spectrum is favored from phenomenological viewpoints of
the SUSY flavor and CP problems \cite{Gabbiani:1996hi} and the
lightest Higgs mass bound \cite{Amsler:2008zzb}. Since it is safe from
the cosmological gravitino over-production problem
\cite{Weinberg:1982zq}, it is also consistent with the thermal
leptogenesis \cite{Fukugita:1986hr}.

In the anomaly mediation the neutral component of $SU(2)_L$ gauginos,
called as Winos, becomes the lightest in gaugino sector.  This is
because the gaugino masses are proportional to the beta functions of
the gauge coupling.  Higgsino, on the other hand, can be as heavy as
gravitino, depending on the K\"ahler potential.  Therefore, the
neutral Wino can be the lightest SUSY particle (LSP) in the anomaly
mediation scenario, and becomes a viable candidate for dark matter in
the universe.

The thermal relic abundance of the Wino LSP in the universe is
consistent with the WMAP observation when the Wino mass is from
2.7~TeV to 3.0~TeV \cite{Hisano:2006nn}.  The lighter Wino predicts
too small thermal relic density; however, it is known that decay of
gravitino or other quasi-stable particles may produce wino
non-thermally so that the relic abundance is consistent with the
observation \cite{Gherghetta:1999sw,Moroi:1999zb}. The successful
Big-Bang Nucleosynthesis (BBN) also gives bounds on the annihilation
cross section of the dark matter, while the large dark matter
annihilation in the BBN era may give a solution to the lithium problem
\cite{Jedamzik,khori}.  In the anomaly mediation, the Wino mass around
(150-300)~GeV may be compatible with the lithium problem when the Wino
LSP is the dominant component of the dark matter.

The direct detection of dark matter is now performed in several
experiments with high sensitivities, and its theoretical sides are
also extensively studied. The tree-level contribution to the Wino
LSP-nucleon ($\tilde{\chi}^0$-$N$) scattering cross section, which is
responsible for direct detection of the dark matter, is evaluated at
Ref.~\cite{Murakami:2000me}. However, in the case that the SUSY
particles and the heavier Higgs bosons have masses of the order of the
gravitino mass except the gauginos in the SUSY SM, the tree-level
interactions of the Wino LSP with quarks are suppressed by the
gravitino mass. Thus, the Wino LSP-nucleon scattering process is
dominated by the weak gauge boson loop diagrams.  However, despite the
loop factor, it was pointed out that the loop contribution is not
suppressed by the Wino mass even if it is heavier than the weak scale
\cite{Hisano:2004pv}.

In this letter, we reevaluate the Wino LSP-nucleon scattering cross
section. The one-loop contribution to the process is evaluated by
Refs.~\cite{Hisano:2004pv, Cirelli:2005uq,Essig:2007az}; however,
their results are not consistent with each other.  In addition, while
the interaction of Wino and gluon is generated by two-loop diagrams,
it has to be included for the complete evaluation of the
spin-independent Wino LSP-nucleon interaction. We take into account
all the relevant diagrams up to two-loop and derive effective
operators, which act as leading contribution in the scattering
process.

\section{Effective interaction  for Wino LSP-nucleon scattering}

First, we summarize the effective interactions of the Wino LSP with light
quarks ($q=u,d,s$) and gluon, which are relevant to the Wino
LSP-nucleon scattering. They are given as follows,
\begin{eqnarray}
{\cal L}^{\rm eff}&=&\sum_{q=u,d,s}{\cal L}^{\rm{eff}}_q +{\cal L}^{\rm{eff}}_g \ ,
\end{eqnarray}
where 
\beq 
{\cal L}^{\rm{eff}}_q
&=& 
d_q\ \oneui\gamma^{\mu}\gamma_5\tilde{\chi}^0\  \bar{q}\gamma_{\mu}\gamma_5 q
+
f_q m_q\ \oneui\tilde{\chi}^0\ \bar{q}q 
+f_q'\ \oneui\tilde{\chi}^0\ \bar{q} i\sla{\partial}q
\cr
&+& \frac{g^{(1)}_q}{m_{\tilde{\chi}^0}} \ \oneui i \partial^{\mu}\gamma^{\nu} 
\tilde{\chi}^0 \ {\cal O}_{\mu\nu}^q
+ \frac{g^{(2)}_q}{m^2_{\tilde{\chi}^0}}\
\oneui(i \partial^{\mu})(i \partial^{\nu})
\tilde{\chi}^0 \ {\cal O}_{\mu\nu}^q \
,
\label{eff_lagq}
\\
{\cal L}^{\rm eff}_{ g}&=&
f_G\ \oneui\tilde{\chi}^0 G_{\mu\nu}^aG^{a\mu\nu}
\nonumber\\
&+&\frac{g^{(1)}_G}{m_{\tilde{\chi}^0}}\
 \oneui i\partial^{\mu}\gamma^{\nu}
\tilde{\chi}^0 \ {\cal O}_{\mu\nu}^g
+
\frac{g^{(2)}_G}{m^2_{\tilde{\chi}^0}}\
\oneui(i\partial^{\mu}) (i\partial^{\nu})\tilde{\chi}^0 
\
{\cal O}_{\mu\nu}^g \ .
\label{eff_lagg}
\eeq
Here, $m_{\tilde{\chi}^0}$ and $m_q$ is mass of Wino and quark,
respectively.  The first term of ${\cal L}^{\rm eff}_q$ contributes to
the spin-dependent $\tilde{\chi}^0$-$N$ interaction, while the other
terms in ${\cal L}^{\rm eff}_{q}$ and ${\cal L}^{\rm eff}_{g}$
generate spin-independent ones. The fourth and fifth terms in ${\cal
  L}^{\rm eff}_q$ and the second and third terms in ${\cal L}^{\rm
  eff}_g$ depend on the twist-2 operators (traceless parts of the
energy momentum tensor) for quarks and gluon,
\beq {\cal O}_{\mu\nu}^q&\equiv&\frac12 \bar{q} i
\left(\partial_{\mu}\gamma_{\nu} +
  \partial_{\nu}\gamma_{\mu} -\frac{1}{2}g_{\mu\nu}\sla{\partial}
\right) q \ ,
\nonumber\\
{\cal O}_{\mu\nu}^g&\equiv&\left(G_{\mu}^{a\rho}G_{\rho\nu}^{a}+
  \frac{1}{4}g_{\mu\nu} G^a_{\alpha\beta}G^{a\alpha\beta}\right) \ .
\eeq

The scattering cross section of the Wino LSP  with  target nuclei is
expressed compactly by using the coefficients given in ${\cal L}^{\rm
eff}_{q}$ and ${\cal L}^{\rm eff}_{g}$ as follows \cite{Jungman:1995df},
\begin{eqnarray}
  \sigma&=&
  \frac{4}{\pi}\left(\frac{m_{\tilde{\chi}^0} m_T}{m_{\tilde{\chi}^0} +m_T}\right)^2
  \left[(n_p f_p+n_nf_n)^2+4 \frac{J+1}{J} \left(
      a_p\left\langle S_p\right\rangle+ a_n\left\langle 
        S_n\right\rangle \right)^2\right]\ , 
\label{sigma}
\end{eqnarray}
where $m_T$ is the mass of target nucleus.  The first term in the
bracket comes from the spin-independent interactions while the second
one is generated by the spin-dependent one.  In the spin-independent
interaction term, $n_p$ and $n_n$ are proton and neutron numbers in
the target nucleus, respectively, and the spin-independent coupling of
the Wino with nucleon, $f_N~(N=p,n)$, is given as
\begin{eqnarray}
f_N/m_N&=&\sum_{q=u,d,s} \left(
(f_q+f_q') f_{Tq}+\frac{3}{4} (q(2)+\bar{q}(2))(g_q^{(1)}+g_q^{(2)})\right)
\nonumber\\
&-&\frac{8\pi}{9\alpha_s}f_{TG} f_G 
+\frac{3}{4} G(2)\left(g^{(1)}_G
+g^{(2)}_G\right) \ . \label{f}
\end{eqnarray}
The matrix elements of nucleon are expressed by using nucleon mass
$m_N$ ($N=p,n$) as\footnote{
  We use equations of motion for quarks for evaluation of the matrix
  elements of $\langle N \vert \bar{q} i\sla{\partial}q \vert
  N\rangle$, though this term is not relevant to our calculation,
  which we will see in the next section.  }
\begin{eqnarray}
f_{Tq}&\equiv& \langle N \vert m_q \bar{q} q \vert N\rangle/m_N \ ,
\nonumber
\\
f_{TG}&\equiv& 1-\sum_{u,d,s}f_{Tq} \ ,
\nonumber\\
\langle N(p)\vert 
{\cal O}_{\mu\nu}^q
\vert N(p) \rangle 
&=&\frac{1}{m_N}
(p_{\mu}p_{\nu}-\frac{1}{4}m^2_N g_{\mu\nu})\
(q(2)+\bar{q}(2)) \ ,
\nonumber\\
\langle N(p) \vert 
{\cal O}_{\mu\nu}^g
\vert N(p) \rangle
& =& \frac{1}{m_N}
(p_{\mu}p_{\nu}-\frac{1}{4}m^2_N g_{\mu\nu})\ 
G(2) \ .
\end{eqnarray}
Here, $q(2)$, $\bar{q}(2)$ and $G(2)$ are the second moments of the
quark, anti-quark and gluon distribution functions, which are
expressed as
\begin{eqnarray}
q(2)+ \bar{q}(2) &=&\int^{1}_{0} dx ~x~ [q(x)+\bar{q}(x)] \ ,
\cr
G(2) &=&\int^{1}_{0} dx ~x ~g(x) \ .
\end{eqnarray}
They are scale-dependent, and are mixed with each others once the QCD
radiative corrections are included. We use the second moments for
gluon and quark distribution functions at the scale of $Z$ boson mass,
which are derived by the CTEQ parton distribution
\cite{Pumplin:2002vw}, and include bottom and charm quark
contributions.  On the other hand, the constant $a_N$ ($N=p,n$), which
is responsible for the spin-dependent contribution, is defined as
\begin{eqnarray}
a_{N}&=&\sum_{q=u,d,s} d_q \Delta q_N \ ,
\end{eqnarray}
\begin{eqnarray}
2 s_{\mu}\Delta q_N &\equiv& \langle N \vert 
\bar{q}\gamma_{\mu}\gamma_5  q \vert N \rangle \ ,
\end{eqnarray}
where $s_{\mu}$ is the nucleon's spin, while $J$ and $\langle
S_N\rangle= \langle A\vert S_N\vert A\rangle$ in Eq.~(\ref{sigma}) are
total spin of nucleus $A$ and the expectation values of the total spin
of protons and neutrons in $A$, respectively.  

\begin{table}
\begin{center}
\begin{tabular}{|l|l|}
\hline
\multicolumn{2}{|c|}{For proton}\cr
\hline
$f_{Tu}$& 0.023\cr
$f_{Td}$& 0.034\cr
$f_{Ts}$&0.025\cr
\hline
\multicolumn{2}{|c|}{For neutron}\cr
\hline
$f_{Tu}$&0.019\cr
$f_{Td}$& 0.041\cr
$f_{Ts}$& 0.025 \cr
\hline
\end{tabular}
\hskip 1cm 
\begin{tabular}{|l|l|}
\hline
\multicolumn{2}{|c|}{Spin fraction}\cr
\hline
$\Delta u$& 0.77\cr
$\Delta d$& -0.49\cr
$\Delta s$& -0.15\cr
\hline 
\end{tabular}
\hskip 1cm 
\begin{tabular}{|l|l||l|l|}
\hline
\multicolumn{4}{|c|}{Second moment at $\mu=m_Z$}\cr
\hline
$G(2)$&0.48&&\cr
$u(2)$&0.22&$\bar{u}(2)$& 0.034\cr
$d(2)$&0.11&$\bar{d}(2)$&0.036\cr
$s(2)$&0.026&$\bar{s}(2)$&0.026\cr
$c(2)$&0.019&$\bar{c}(2)$&0.019\cr
$b(2)$&0.012&$\bar{b}(2)$&0.012\cr
\hline
\end{tabular}
\end{center}
\caption{   Parameters for quark and gluon matrix elements used in this letter.  $f_{Ti}$ $(i=u,d,s)$ is taken from the estimation in Refs.~\cite{Cheng:1988im,Ohki:2008ff}. The spin fractions for proton  comes from  Ref.~\cite{Adams:1995ufa}. Those for neutron are given by exchange of up and down quarks in the tables.  The second moments for
  gluon and quark distribution functions are calculated at the scale $\mu=m_Z$ ($m_Z$ is $Z$ boson mass) using the CTEQ parton distribution \cite{Pumplin:2002vw}. 
}
\label{table1}
\end{table}

Notice that the term proportional to $f_G$ in the spin-independent
coupling of $\tilde{\chi}^0$-$N$ in Eq.~(\ref{f}) is divided by
$\alpha_s$.  It comes from definition of the gluon contribution to
nucleon mass, $f_{TG}$, and the trace anomaly of the energy momentum
tensor as
\begin{eqnarray}
m_N f_{TG}&=&
-\frac{9\alpha_s}{8\pi } \langle N \vert G_{\mu\nu}^aG^{a\mu\nu} \vert N\rangle
\end{eqnarray}
at the leading order.\footnote{Here, we use three-flavor
  approximation.}  Thus, when evaluating the spin-independent
$\tilde{\chi}^0$-$N$ interaction, we need to include $O(\alpha_s)$
correction to $f_G$ \cite{Drees:1993bu}. Other contributions in the Wino
LSP and gluon interaction, which come from gluon twist-2 operators, are
sub-leading as far as the coefficients are $O(\alpha_s)$. Thus, we
neglect the contribution from gluon twist-2 operators in the following
discussion.

Parameters for quark and gluon matrix elements used in this analysis
are summarized in Table~1. Notice that the strange quark content of
the nucleon $f_{Ts}$ is much smaller than previous thought according
to the recent lattice simulation \cite{Ohki:2008ff}. This leads to
significant suppression on the spin-independent cross section, then
the interaction of Wino and gluon becomes relatively more important in
the cross section.

\section{Results}

Now we evaluate the coefficients of effective interactions in
Eqs.~(\ref{eff_lagq}, \ref{eff_lagg}), which are needed to calculate
the scattering cross section.

The Wino LSP accompanies the charged Wino ($\tilde{\chi}^-$). The mass
difference is dominated by one-loop contribution unless Higgsino and
Wino masses are almost degenerate; we ignore it in this letter.  The
coupling of neutral and charged Winos to the standard model sector is
only through gauge interactions as
\begin{eqnarray}
  {\cal L}_{\rm int}
  &=&
  -\frac{e}{s_W}
  \left(
    \overline{\tilde{\chi}^0}\gamma^\mu\tilde{\chi}^-W^\dagger_\mu
    +
    h.c.
  \right)
  +
  e \frac{c_W}{s_W}\overline{\tilde{\chi}^-}\gamma^\mu\tilde{\chi}^-Z_\mu
  +
  e\overline{\tilde{\chi}^-}\gamma^\mu\tilde{\chi}^-A_\mu \ .
\end{eqnarray}
Here, $e$ is the electric charge, $s_W=\sin\theta_W$ and
$c_W=\cos\theta_W$ with $\theta_W$ being the Weinberg angle.  As is
described in Introduction, the effective interactions of the Wino LSP
to light quarks are generated by the loop diagrams.  The leading
contribution comes from one-loop interaction, which is shown in
Fig.~1.  After calculating the diagrams, the coefficients in
Eq.~(\ref{eff_lagq}) are derived as follows,
\begin{eqnarray}
f_q &=&\frac{\alpha_2^2}{4m_W m_{h^0}^2} g_{\rm H}(x) \ ,
\label{fq} \\
f_q' &=& 0 \ ,\nonumber\\
d_q&=& \frac{\alpha_2^2}{m_W^2} g_{\rm AV}(x) \ ,\\
g_q^{(1)}&=&\frac{\alpha_2^2}{m_W^3} g_{T1}(x) \ ,\\
g_q^{(2)}&=&\frac{\alpha_2^2}{m_W^3} g_{T2}(x) \ ,
\label{1loop}
\end{eqnarray}
where $m_{h^0}$ is the lightest Higgs boson ({\it i.e.}, SM Higgs
boson) mass, $x=m_W^2/m^2_{\tilde{\chi}^0}$ and
$\alpha_2=\alpha/s_W^2$ (here $m_W$ is the $W$ boson mass and $\alpha$
is the fine-structure constant). The diagram (a) in Fig.~1, which is
induced by the SM Higgs boson $(h^0)$ exchange, contributes to $f_q$,
while the diagram (b) generates the other terms in
Eq.~(\ref{eff_lagq}).  With the light quark masses ignored, the mass
functions in Eqs.~(\ref{fq}-\ref{1loop}) are given as
\begin{eqnarray}
g_{\rm H}(x)
&=&
-\frac{2}{b}
(2+2x-x^2)\tan^{-1}(\frac{2 b}{\sqrt{x}})
+
2\sqrt{x}(2-x \log(x)) \ ,
\nonumber\\
g_{\rm AV}(x)
&=&
\frac{1}{24b}\sqrt{x}(8-x-x^2)\tan^{-1}(\frac{2b}{\sqrt{x}})
-\frac{1}{24} x(2-(3+x)\log(x)) \ ,
\label{gav}
\nonumber\\
g_{\rm T1}(x)
&=&
\frac{1}{3}b(2+x^2)\tan^{-1}(\frac{2 b}{\sqrt{x}})
+\frac{1}{12}\sqrt{x}(1-2x-x(2-x)\log{x}) \ , 
\nonumber\\
g_{\rm T2}(x)
&=&
\frac{1}{4b} x (2-4x+x^2)\tan^{-1}(\frac{2 b}{\sqrt{x}})
-\frac{1}{4}\sqrt{x}(1-2x-x(2-x)\log(x)) \ ,
\nonumber\\
\end{eqnarray}
with ${b}=\sqrt{1-x/4}$.\footnote
{Here, $g_{T1}$ is larger than $F^{(0)}_{T1}$ given in Eq. (42) in
  \cite{Hisano:2004pv}.  We corrected it in this calculation.  }
%
\begin{figure}[t]
 \begin{center}
   \includegraphics[width=0.55\linewidth]{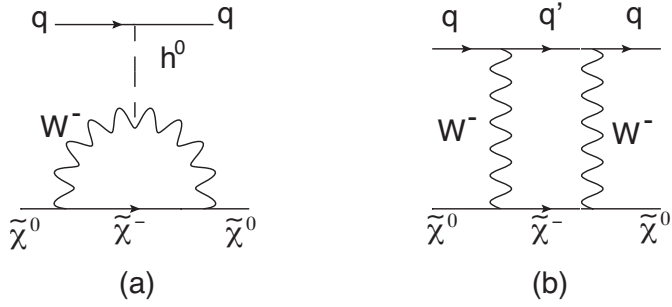} 
   \caption{One-loop contributions to effective interactions of
     Wino LSP and light quarks.}
 \end{center}
\end{figure}
As discussed in Ref.~\cite{Hisano:2004pv}, the spin-independent
interaction of $\tilde{\chi}^0$-$N$ are not suppressed even if the
Wino LSP is much larger than the $W$ boson mass. The mass functions
$g_{\rm H}(x)$ and $g_{\rm T1}(x)$ become finite in a limit of
$x\rightarrow 0$ while other two functions are zero, as 
\begin{eqnarray}
g_{\rm H}(x)&\simeq& -2\pi \ ,
\nonumber\\
g_{\rm AV}(x)&\simeq& \frac{\sqrt{x}}6 \pi \ ,
\nonumber\\
g_{\rm T1}(x)&\simeq& \frac{\pi}3 \ ,
\nonumber\\
g_{\rm T2}(x)&\simeq&  -\frac{\sqrt{x}}6 \ .
\end{eqnarray}
%

\begin{figure}[t]
 \begin{center}
   \includegraphics[width=0.7\linewidth]{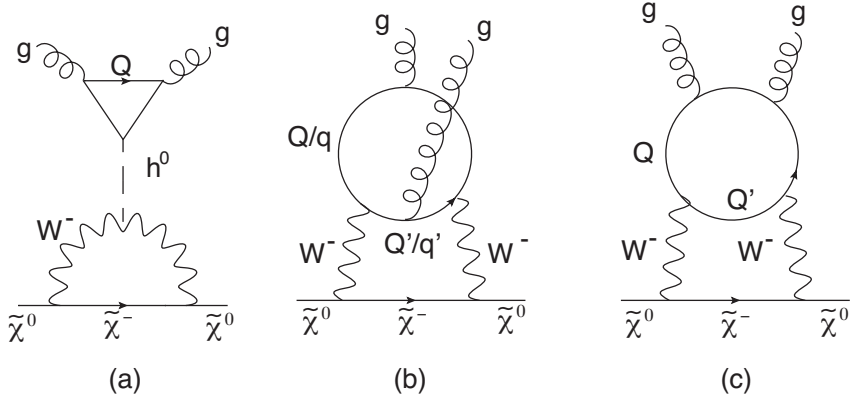} 
   \caption{Two-loop contributions to interactions of Wino LSP and
     gluon. Here, $Q$ and $q$ represent heavy and light quarks,
     respectively.  }
 \end{center}
\end{figure}
Next, let us discuss the effective interactions of the Wino LSP and
gluon. As we discussed in the previous section, the $O(\alpha_s)$
correction to $f_G$ in Eq.~(\ref{eff_lagg}) is relevant at the leading
order though it is induced by two-loop order.  Three types of diagrams
in Fig.~2 contribute to $f_G$.  The diagram (a) includes heavy quark
loop ($Q=c,b,t$). The heavy quark content of the nucleon is related to
the gluon condensate as \cite{Shifman:1978zn}
\begin{eqnarray}
\langle N \vert m_Q \bar{Q}Q \vert N\rangle
&=&-\frac{\alpha_s}{12\pi}
\langle N \vert G_{\mu\nu}^aG^{a\mu\nu} \vert N\rangle \ .
\label{shifman}
\end{eqnarray}
Thus, the diagram (a) can be evaluated from Eq.~(\ref{fq}) by
replacing light to heavy quarks and using Eq.~(\ref{shifman}).  On the
other hand, we need to calculate irreducible two-loop diagrams (b) and
(c) explicitly.  In the diagram (c), the momentum which dominates the
quark loop integration is characterized by mass of quark which emits
two gluons.  Since we are constructing the effective theory under
$O(1)$ GeV, the integration in the infrared regime under such energy
scale should not be included.  Thus, light quarks does not contribute
in this diagram.  On the other hand, the loop momentum of quark loop
in the diagram (b) is dominated by the external momentum of the quark
loop diagram ({\it i.e.}, $W$ boson mass in this case); therefore, all
quarks contributes in the loop.
We express the $O(\alpha_s)$ contribution to $f_G$ as follows,
\begin{eqnarray}
f_G &=& -3\times \frac{\alpha_s}{12\pi} 
\frac{\alpha_2^2}{4m_W m_{h^0}^2} g_{\rm H}(x)
+\frac{\alpha_s}{4\pi} 
\frac{\alpha_2^2}{m_W^3} g_{\rm B3}(x,y)
+2\times \frac{\alpha_s}{4\pi} 
 \frac{\alpha_2^2}{m_W^3} g_{\rm B1}(x) \ ,
\end{eqnarray}
where $y=m_t^2/m^2_{\tilde{\chi}^0}$ ($m_t$ is top quark mass).  The
first term represents the contribution from the diagram (a). The
second term comes from the diagrams (b) and (c) with the
third-generation quark loop, while the the first- and
second-generation quarks contribute to the third one. We ignore quark
masses except for the top quark. Then, we found that the mass
functions $g_{\rm B3}(x,y)$ and $g_{\rm B1}(x)$ are given by
\begin{eqnarray}
g_{\rm B3}(x,y)&=&
-\frac{x^{3/2} (2 y-x)}{12
   (y-x)^2}
-\frac{x^{3/2} y^3 \log (y) }{24
   (y-x)^3}
+\frac{x^{5/2} (3 y^2-3
   x y+x^2 ) \log (x)}{24
   (y-x)^3}
\nonumber\\
&+&\frac{x^{3/2} \sqrt{y} (y^3-2
   y^2-14 y+6 x) \tan^{-1}(\frac{2 b_t}{\sqrt{y}}) }{24
   b_t (y-x)^3 }
\nonumber\\
&-&\frac{x \left(x^4-3 y
   x^3-2 x^3+3 y^2 x^2+6 y
   x^2+4 x^2-6 y^2 x-6 y
   x-6 y^2\right) \tan
   ^{-1}(\frac{2 b}{\sqrt{x}})}{24 b
   (y-x)^3} \ ,
\nonumber\\ 
g_{\rm B1}(x)&=&
-\frac{1}{24} \sqrt{x} (x \log
   (x)-2)
+\frac{(x^2-2x
   +4) \tan ^{-1}(\frac{2 b}{\sqrt{x}})}{24b} \ ,
\end{eqnarray}
where $b_t=\sqrt{1-y/4}$. Notice that the diagrams (b) and (c) also
give finite contributions to the spin-independent $\tilde{\chi}^0$-$N$
interaction in a limit of $m_{\tilde{\chi}^0}\rightarrow \infty$,
{\it i.e.}, $x,~y\ll 1$,
\begin{eqnarray}
g_{\rm B3}(x,y)&\simeq&\frac{(3\sqrt{y}+2\sqrt{x}) x} 
{24(\sqrt{x}+\sqrt{y})^3} \pi \ ,\nonumber\\
g_{\rm B1}(x)&\simeq& \frac{\pi}{12} \ .
\end{eqnarray}

Now we are at the position to present the scattering cross section.
In Fig.~3, we show the spin-independent $\tilde{\chi}^0$-$p$
scattering cross section as a function of $m_{\tilde{\chi}^0}$ (solid
line). Here, we take $m_{h^0}= 115$, $130$, 300~GeV, and 1~TeV from
bottom to top. While the latter two values may not be realistic in the
minimal SUSY SM, the next-minimal SUSY SM (NMSSM), for example, may
predict larger Higgs boson mass. It was found that the
spin-independent cross section is $O(10^{-(48-46)})$ cm$^2$, depending
on the Higgs boson mass.

In order to understand the result, we also plot each contribution from
the effective operators in $f_p$ in Fig.~4.  Solid line represents the
Higgs exchange contribution (Fig.~1(a) and Fig.~2(a)), dashed line is
for the twist-2 operator contribution (Fig.~1(b)), and dash-dot line
is for that from irreducible two-loop diagrams in Fig.~2(b) and
(c). As is seen, the contribution from quark twist-2 operators is
dominant part.  However, we also found that other two also give
relatively large contribution by the opposite sign.  Consequently,
$f_p$ is suppressed by the accidental cancellation, which leads to the
smaller spin-independent cross section.  When the Higgs boson mass is
taken to be larger, the cross section becomes larger since the
cancellation is milder.

In this letter, we have ignored the tree-level coupling of the Wino
LSP and the lightest Higgs boson since it is suppressed by heavy
Higgsino mass. When it dominates the spin-independent interaction, the
spin-independent cross section is evaluated as
\begin{eqnarray}
\sigma_p&\simeq&9\times 10^{-47} {\rm cm^2} \times
\left(\frac{m_{\tilde{H}}}{10{\rm TeV}}\right)^{-2}
\left(\frac{m_{h^0}}{115{\rm GeV}}\right)^{-4}
\sin^22\beta \ ,
\end{eqnarray}
where $m_{\tilde{H}}$ is the Higgsino mass and $\beta$ is the vacuum
angle in the SUSY SM. Thus, the tree-level contribution may dominate
the spin-independent cross section, depending on parameters in the
SUSY SM, even if the the SUSY particle masses are of the order of the
gravitino mass.  

\begin{figure}[t]
 \begin{center}
   \includegraphics[width=0.6\linewidth]{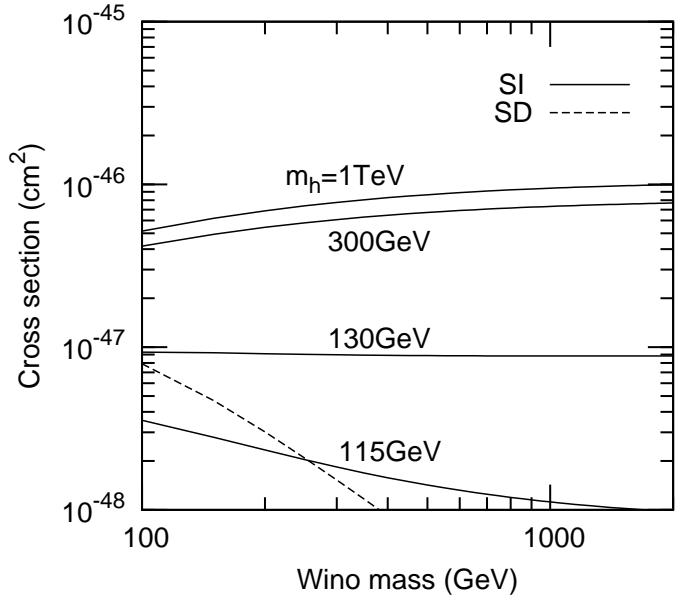} 
   \caption{$\tilde{\chi}^0$-$p$ scattering cross section as a
     function of $m_{\tilde{\chi}^0}$. Spin-independent (SI) cross
     section is given in solid line, taking $m_{h^0}=115,~130~{\rm
       GeV}$, 300~GeV, and 1~TeV from bottom to top.  Here, we also
     plot spin-dependent (SD) one in dashed line. }
 \end{center}
\label{fig:sigma}
\end{figure}

For completeness, in Fig.~\ref{fig:sigma}, we also show the
spin-dependent cross section in dashed line. As expected from the
behavior of the mass function $g_{\rm AV}(x)$, it was found that the
cross section is suppressed by the Wino mass.

Finally, we comment difference between our result and the previous
works.  We found a few errors in the calculation in
Ref.~\cite{Hisano:2004pv} as described in this text, though the
qualitative behavior is not different from this work. On the other
hand, compared with Refs.~\cite{Cirelli:2005uq,Essig:2007az}, our
result for the spin-independent cross section is smaller by
$O(10^{-(2-3)})$.  In Ref.~\cite{Cirelli:2005uq} only the contribution
to scalar couplings to quarks and gluon are evaluated.  In
Ref.~\cite{Essig:2007az} the relative sign of the quark twist-2 and
the Higgs boson exchange contributions is opposite to ours. Thus, the
cross section is not reduced by accidental cancellation in those
works. Their loop functions are also different from ours. We could not
understand origin of the differences.

\begin{figure}[t]
 \begin{center}
   \includegraphics[width=0.6\linewidth]{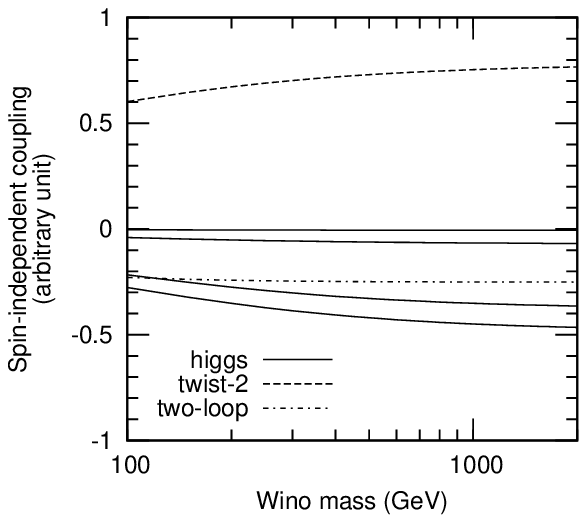} 
   \caption{Each contribution in spin-independent coupling, $f_p$.  Solid
     line represents the Higgs exchange contribution including heavy
     quark one, dashed line is for the twist-2 operator contribution,
     and dash-dot line is for that from two-loop diagrams in Fig.~1(b)
     and (c).  The Higgs boson mass is $m_{h^0}=115,~130~{\rm GeV}$,
     300~GeV, and 1~TeV from bottom to top.  Here, unit is
     arbitrary. }
 \end{center}
\label{fig:fp}
\end{figure}

\section{Conclusion and discussion}

In this letter, we calculated the Wino LSP-nucleon cross section in the
anomaly-mediated SUSY breaking mechanism.  We especially consider the
scenario in which all SUSY particles except for gauginos are heavy to
decouple in electroweak scale, and neutral Wino becomes the LSP.  In
such a scenario, although the Wino LSP does not interact with nucleon at
tree-level, it does in loop diagrams.  We have taken into account all
the loop diagrams which act as leading contributions to the Wino
LSP-nucleon scattering.  As a result, the spin-independent cross
section turns out to be $O(10^{-(48-46)})~{\rm cm}^2$, depending on
the Higgs boson mass.  In the calculation, we found that Wino-gluon
interaction at two-loop level contributes in opposite sign to the main
Wino-quark interaction, which leads to the suppression of the total
Wino-nucleon coupling.  Therefore, it is concluded that the direct
detection of Wino dark matter is difficult in the present experiments
in the scenario.

We comment on the cancellation that we observed in Wino-nucleon
coupling.  The cancellation is supposed to be accidental in the
scenario that we analysed.  Thus, if one consider the other scenarios
in SUSY or other models, the two-loop may contribute as large as the
lower order diagrams, which may cause enhancement of scattering cross
section.  Such analysis will be given elsewhere
\cite{HisanoIshiwataNagata}.



\section*{Acknowledgment}

The work was supported in part by the Grant-in-Aid for the Ministry of
Education, Culture, Sports, Science, and Technology, Government of
Japan, No. 20244037, No. 2054252 and No. 2244021 (J.H.) and Research
Fellowships of the Japan Society for the Promotion of Science for
Young Scientists (K.I.).  The work of J.H. is also supported by the
World Premier International Research Center Initiative (WPI
Initiative), MEXT, Japan.


{}

\end{document}